\documentclass[12pt,a4paper,final]{article}

\usepackage{a4wide}
\usepackage{amssymb}
\usepackage[english]{babel}
\usepackage{color}
\usepackage[colorlinks=false,%
            pdfkeywords={OCaml, Just In Time compilation, byte-code, machine code generation},%
            pdftitle={Just-In-Time compilation of OCaml byte-code},%
            pdfauthor={Benedikt Meurer},%
            pdfsubject={},%
            pdfdisplaydoctitle=true]{hyperref}
\usepackage{tikz}
\usepackage{varwidth}

\usetikzlibrary{arrows}
\usetikzlibrary{trees}

\definecolor{core2byt}{HTML}{7FFFD4}
\definecolor{core2jit2}{HTML}{CD5C5C}
\definecolor{xeonbyt}{HTML}{2F4F4F}
\definecolor{xeonjit2}{HTML}{9932CC}
\definecolor{p4byt}{HTML}{32CD32}
\definecolor{p4jit}{HTML}{FFD700}
\definecolor{p4jit2}{HTML}{4169E1}

\begin{document}

\title{%
  Just-In-Time compilation of OCaml byte-code
}
\author{%
  Benedikt Meurer\\
  Compilerbau und Softwareanalyse\\
  Universit\"at Siegen\\
  D-57068 Siegen, Germany\\
  \url{meurer@informatik.uni-siegen.de}
}
\date{}
\maketitle
\begin{abstract}
  This paper presents various improvements that were applied to
  \textsc{OCamlJit2}, a Just-In-Time compiler for the \textsc{OCaml}
  byte-code virtual machine. \textsc{OCamlJit2} currently runs on
  various Unix-like systems with x86 or x86-64 processors. The
  improvements, including the new x86 port, are described in detail,
  and performance measures are given, including a direct
  comparison of \textsc{OCamlJit2} to \textsc{OCamlJit}.
\end{abstract}

\section{Introduction}

The \textsc{OCaml} system \cite{Leroy10,Remy02} is the main implementation of the
\textsc{Caml} language \cite{Caml10}, featuring a powerful module system combined with a full-fledged
object-oriented layer. It comes with an optimizing native code compiler \texttt{ocamlopt},
a byte-code compiler \texttt{ocamlc} with an associated runtime \texttt{ocamlrun}, and
an interactive top-level \texttt{ocaml}.
\textsc{OCamlJit} \cite{Starynkevitch04} and \textsc{OCamlJit2} \cite{Meurer10ocamljit}
provide Just-In-Time compilers for the byte-code used by \texttt{ocamlrun} and
\texttt{ocaml}.

We describe a set of improvements that were applied to \textsc{OCamlJit2},
including a new port to the x86 architecture\footnote{Also known als i386 or IA-32 architecture,
but we prefer the vendor-neutral term.} and some interesting optimizations to further
improve the execution speed of the JIT engine. \textsc{OCamlJit2} is open-source\footnote{The
full source code is available from \url{https://github.com/bmeurer/camljit2/} under the terms of
the Q Public License and the GNU Library General Public License.} and is verified to work with
Linux/amd64, Linux/i386 and Mac OS X (64-bit Intel Macs only), but should also run on other
Unix-like systems with x86 or x86-64 processors.

The paper is organized as follows: Section~\ref{section:Existing_systems} mentions existing
systems and relevant sources of information. Section~\ref{section:Improvements} describes
the improvements applied to \textsc{OCamlJit2}, in particular the new x86 port and various
floating-point optimizations. Detailed performance measures are given in
section~\ref{section:Performance}, including a direct comparison with
\textsc{OCamlJit}. Section~\ref{section:Conclusions_and_further_work} concludes with
possible directions for further work.

\section{Existing systems} \label{section:Existing_systems}

The implementation of \textsc{OCamlJit2} described below is based on a previous
prototype of \textsc{OCamlJit2} \cite{Meurer10ocamljit}, and earlier work on
\textsc{OCamlJit} \cite{Starynkevitch04} and \textsc{OCaml} \cite{Leroy90,Leroy10,Leroy02,Remy02}.
We assume that the reader is familiar with the internals of the \textsc{OCaml} byte-code
and runtime, and the structure of the \textsc{OCamlJit} and \textsc{OCamlJit2} Just-In-Time
compilers. An overview of the relevant parts of the \textsc{OCaml} virtual machine is given
in both \cite{Meurer10ocamljit} and \cite{Starynkevitch04}, while \cite{Remy02} provides
the necessary insights on the \textsc{OCaml} language.

\section{Improvements} \label{section:Improvements}

This section describes the improvements that were applied to the \textsc{OCamlJit2} prototype
described in \cite{Meurer10ocamljit}. This
includes the new x86 port (section~\ref{subsection:32_bit_support}), a simple native code managment
strategy (section~\ref{subsection:Native_code_management}), as well as improvements
in the implementation of floating-point primitives (section~\ref{subsection:Floating_point_optimizations}),
which reduced the execution time of floating-point benchmarks by up to $30\%$.
Readers only interested in the performance results may skip straight to section~\ref{section:Performance}.

\subsection{32-bit support} \label{subsection:32_bit_support}

\textsc{OCamlJit2} now supports both x86-64 processors (operating in long mode)
as well as x86 processors (operating in protected mode). This section provides a brief
overview of the mapping of the \textsc{OCaml} virtual machine to the x86
architecture, especially the mapping of the virtual machine registers to the available
physical registers. See \cite{Meurer10ocamljit} for a description of the implementation
details of the x86-64 port.

The x86 architecture \cite{Intel10Vol1} provides $8$ general purpose 32-bit registers
\texttt{\%eax}, \texttt{\%ebx}, \texttt{\%ecx}, \texttt{\%edx}, \texttt{\%ebp}, \texttt{\%esp},
\texttt{\%edi} and \texttt{\%esi}, as well as $8$ 80-bit floating-point registers organized
into a so-called \emph{FPU stack} (and also used as 64-bit MMX registers). Recent incarnations also
include a set of $8$ SSE2 registers \texttt{\%xmm0}, $\ldots$, \texttt{\%xmm7}.
The System V ABI for the x86 architecture \cite{SCO97Abi386}, implemented by almost all
operating systems running on x86 processors, except Win32 which uses a different ABI,
mandates that registers \texttt{\%ebp}, \texttt{\%ebx}, \texttt{\%edi}, \texttt{\%esi}
and \texttt{\%esp} belong to the caller and the callee is required to preserve their
values across function calls. The remaining registers belong to the callee.

To share as much code as possible with the x86-64 port, we use a similar register assignment
for the x86 architecture. This includes making good use of callee-save registers to avoid
saving and restoring too many aspects of the machine state whenever a C function is
invoked. Our register assignment therefore looks as follows:

The virtual register \texttt{accu} is mapped to \texttt{\%eax}.
\texttt{extra\_args} goes into \texttt{\%ebx}, which -- just like on x86-64 -- contains
the number of extra arguments as tagged integer.
The environment pointer \texttt{env} goes into \texttt{\%ebp}, and
the stack pointer \texttt{sp} goes into \texttt{\%esi}.
\texttt{\%edi} contains the cached value of the minor heap allocation pointer
\texttt{caml\_young\_ptr} to speed up allocations of blocks in the minor heap; this
is different from \texttt{ocamlopt} on x86, where \texttt{caml\_young\_ptr} is not
cached in a register (unlike x86-64 where \texttt{ocamlopt} caches its value in \texttt{\%r15}).

The setup and helper routines that form the runtime of the JIT engine are located in the
file \texttt{byterun/jit\_rt\_i386.S}. These are mostly copies of their x86-64 counterparts
in the file \texttt{byterun/jit\_rt\_amd64.S}, adapted to the x86 architecture. The
adaption was mostly straight-forward, replacing the x86-64 registers with their x86
counterparts and the 64-bit opcodes with their 32-bit counterparts.

\subsubsection{Address mapping and on-demand compilation}

We use the existing scheme \cite{Meurer10ocamljit} to map byte-code to native machine code address, which
replaces the instruction opcode word within the byte-code sequence with
the offset of the generated native code relative to the base address
\texttt{caml\_jit\_code\_end}. Whenever jumping to a byte-code address - i.e.
during closure application or return -- this offset is read from the instruction
opcode, \texttt{caml\_jit\_code\_end} is added to it, and a jump to the resulting
address is performed. The trampoline code for x86 -- shown in
Figure~\ref{figure:Byte_code_trampoline_x86} -- is therefore quite similar to
the trampoline code for x86-64 (\texttt{\%ecx} contains the address of the
byte-code to execute and \texttt{\%edx} is a temporary register).

\begin{figure}[h]
  \centering
  \begin{varwidth}{\linewidth}
  \begin{verbatim}
movl (%ecx), %edx
addl caml_jit_code_end, %edx
jmpl *%edx
\end{verbatim}
  \end{varwidth}
  \caption{Byte-code trampoline (x86)}
  \label{figure:Byte_code_trampoline_x86}
\end{figure}

Adapting the on-demand compilation driver was also straight-forward, due to
the similarities of the x86 and x86-64 architectures. It was merely a matter
of adding a x86 byte-code compile trampoline -- shown in
Figure~\ref{figure:Byte_code_compile_trampoline_x86} -- which is slightly longer
than its x86-64 counterpart, because the C calling convention \cite{SCO97Abi386}
requires all parameters to be passed on the C stack.

\begin{figure}[h]
  \centering
  \begin{varwidth}{\linewidth}
  \begin{verbatim}
pushl %eax
pushl %ecx
call  caml_jit_compile
movl  %eax, %edx
popl  %ecx
popl  %eax
jmpl  *%edx
\end{verbatim}
  \end{varwidth}
  \caption{Byte-code compile trampoline (x86)}
  \label{figure:Byte_code_compile_trampoline_x86}
\end{figure}

Whenever the byte-code compile trampoline is invoked, \texttt{\%eax} contains
the current \texttt{accu} value, \texttt{\%ecx} contains the byte-code address
and the remainder of the \textsc{OCaml} virtual machine state is located in
global memory locations and callee-save registers. Therefore \texttt{\%eax}
has to be preserved on the C stack and \texttt{\%ecx} must be passed as parameter
to the \texttt{caml\_jit\_compile} function. Upon return \texttt{\%eax} contains
the address of the generated native code, which is saved to \texttt{\%edx}.
Afterwards the stack frame is removed, restoring \texttt{\%eax}, and execution
continues at the native code address.

The remaining porting effort was mostly related to generalizing the code
emitter preprocessor macros to conditionally emit x86 or x86-64 code, and fiddling with
some nasty details of the different addressing modes. The core of the code
generator is almost free of \texttt{\#ifdef}'s, because the structure of the generated code
for the two targets is pretty much equivalent. This is especially true for
floating-point operations, which use SSE2 registers and instructions on
both x86 and x86-64. \textsc{OCamlJit2} therefore requires a processor
with support for the SSE2 extension \cite{Amd09Vol1,Intel10Vol1}; if \textsc{OCamlJit2} detects a
processor without SSE2, it disables the JIT engine and falls back to
the byte-code interpreter\footnote{This is only relevant in case of x86,
as all x86-64 processors are required to implement the SSE2 extension.}.

\subsection{Native code management} \label{subsection:Native_code_management}

With our earlier prototype, memory allocated to a chunk of generated native machine code was never freed
by \textsc{OCamlJit2}, even if the byte-code segment, for which this native machine code
was generated, was released (using the \texttt{Meta.static\_release\_bytecode} OCaml
function). This may cause trouble for MetaOCaml \cite{Taha06} and long-lived interactive
top-level session, where the life-time of most byte-code segments is limited.
This was mostly due to the fact that all generated native machine code was stored 
incrementally in one large memory region, with no way to figure out which part of
the memory region contains code for a certain byte-code segment. We have addressed
this problem with a solution quite similar to that used in \textsc{OCamlJit}
\cite{Starynkevitch04}.

\begin{figure}[htb]
  \centering
  \begin{tikzpicture}
    \filldraw [fill=blue!10] (0,1) rectangle (1,3);
    \filldraw [fill=blue!10] (0,4) rectangle (1,6);
    \filldraw [fill=green!10] (3,1) rectangle (6,6);
    \draw (0.5,6.2) node {\footnotesize $\mbox{segment}_1$};
    \draw (0.5,3.2) node {\footnotesize $\mbox{segment}_2$};
    \draw (4.5,6.2) node {\footnotesize $\mbox{native code area}$};
    \draw (4,2.7) node {\footnotesize $\mbox{chunk}_3$};
    \draw (4,4.2) node {\footnotesize $\mbox{chunk}_2$};
    \draw (4,5.7) node {\footnotesize $\mbox{chunk}_1$};
    \filldraw [fill=green!80] (3.5,1.5) rectangle (4.5,2.5);
    \filldraw [fill=green!80] (3.5,3.0) rectangle (4.5,4.0);
    \filldraw [fill=green!80] (3.5,4.5) rectangle (4.5,5.5);
    \draw [densely dotted,->] (1,2) -- (3.45,2);
    \draw [densely dotted,->] (1,5) -- (3.45,5);
    \draw [densely dotted,->] (4.5,5) .. controls (5.0,4.7) and (5.0,3.7) .. (4.5,3.5);
  \end{tikzpicture}
  \caption{Byte-code segments and native code chunks}
  \label{figure:Byte_code_segments_and_native_code_chunks}
\end{figure}
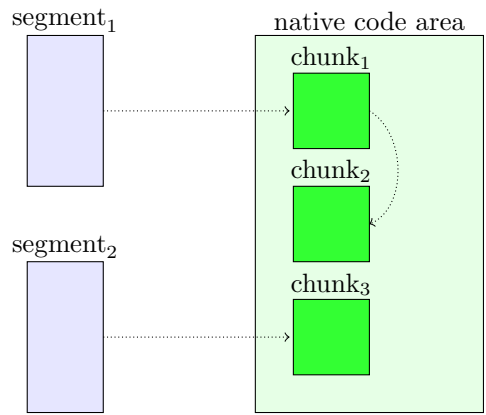

Instead of generating native machine code incrementally into one large region of
memory, we divide the region into smaller parts, called native code \emph{chunks},
and generate code to these chunks. Every byte-code segment has an associated list of
chunks, allocated from the large memory region, now called \emph{chunk pool},
which contain the generated code for the given byte-code segment, as shown
in Figure~\ref{figure:Byte_code_segments_and_native_code_chunks}.

Every segment starts out with an empty list of chunks. As soon as on-demand compilation
triggers for the segment, a new chunk is allocated from the chunk pool and native
machine code is generated into this chunk until it is filled up; when this happens,
the code generator allocates another chunk for this segment, and emits a \texttt{jmp}
instruction from the previous chunk to the new chunk if necessary. Once a byte-code
segment is freed using \texttt{Meta.static\_release\_bytecode}, all associated native code
chunks are also released. This way \textsc{OCamlJit2} no longer leaks
native code generated for previously released byte-code segments.

While this technique is both simple and effective, there are also several drawbacks.
The speed of both the JIT compiler and the generated code is somewhat dependent
on the size of the native code chunks. A small chunk size helps to reduce the size of
the working set and the amount of memory wasted in long-lived interactive top-level
sessions, but on the other hand decreases the throughput of the JIT compilation
driver and leads to somewhat less efficient execution for common byte-code programs
which use only a single byte-code segment. We have settled on a chunk size of $256$ KiB for
now, which seems to provide a good compromise.

A possible way to reduce the amount of wasted memory with the interactive
top-level would be to store code generated for small byte-code segments to some special,
shared code chunks, and manage the life-time of these shared chunks using a
reference counting scheme.

\subsection{Floating-point optimizations} \label{subsection:Floating_point_optimizations}

\textsc{OCaml} uses a boxed representation for floating-point values. It does this
for various reasons, i.e.~to simplify the interface to the garbage collector
and the garbage collector itself. While this is an elegant and portable solution, it decreases
the performance of \textsc{OCaml} programs using floating-point calculations, especially
when executed with the byte-code interpreter.

The optimizing native code compiler \texttt{ocamlopt} applies various optimizations to avoid generating
a boxed floating-point object in the heap for each and every floating-point value during
the execution of the program. The byte-code interpreter \texttt{ocamlrun} however has to box every floating-point
value, which is certainly slower than using the available floating-point registers, and
also causes a non-negligible load on the garbage collector. Both \textsc{OCamlJit} and
\textsc{OCamlJit2} (as described in \cite{Meurer10ocamljit}) used the same strategy
as the byte-code interpreter, namely allocating a heap object for every floating-point
value during the execution, but both JIT engines applied various peephole optimizations
to avoid the overhead of calling the floating-point C primitives.

We have implemented a new technique for \textsc{OCamlJit2}, which avoids the heap
allocation for temporary floating-point values that appear as result of a byte-code
instruction and are used as argument in the subsequent byte-code instruction.
Figure~\ref{figure:Subsequent_floating_point_primitives_almabench} shows an example
taken from the byte-code of the \texttt{almabench.ml} \textsc{OCaml} program.

\begin{figure}[h]
  \centering
  \begin{varwidth}{\linewidth}
\begin{verbatim}
ccall caml_array_unsafe_get_float, 2
ccall caml_mul_float, 2
ccall caml_add_float, 2
ccall caml_add_float, 2
ccall caml_sqrt_float, 1
push
\end{verbatim}
  \end{varwidth}
  \caption{Subsequent floating-point primitives (\texttt{almabench.ml})}
  \label{figure:Subsequent_floating_point_primitives_almabench}
\end{figure}

Executing this piece of byte-code with the byte-code interpreter allocates five
floating-point objects in the minor heap, one for each \texttt{ccall}. The first
four objects will die immediately once the garbage collector is run, since they
are only used as temporary results, while the result of the \texttt{caml\_sqrt\_float}
call may indeed survive for a longer period of time. Furthermore, accessing the
actual floating-point values of the temporary results requires at least four
additional load instructions.
If the temporary results would be kept in registers, there would be no need for
the heap allocation and this would also eliminate the additional load instructions.
While the heap allocations can seriously decrease performance, the additional
load instructions are a minor issue, since the \emph{store-to-load forwarding}
techniques \cite{Austin95,Lipasti96} implemented in modern processors will usually
eliminate the load from memory.

We have implemented a clever optimization in \textsc{OCamlJit2}, which translates
various floating-point primitives using SSE2 instructions and functions from the
standard C math library, in a way that the result of each primitive is not stored in
a heap-allocated object, but is left in the \texttt{\%xmm0} register. Subsequent
floating-point primitives then take the value from the \texttt{\%xmm0} register
instead of the memory location pointed to by \texttt{\%rax} (or \texttt{\%eax}).
This process is repeated for all floating-point primitives in a row. The last
floating-point instruction -- the call to \texttt{caml\_sqrt\_float} in our
example -- then allocates a heap block for its result.
Our optimization is particularly beneficial for the x86 port, where we were able
to beat the optimizing native code compiler in the \texttt{almabench.unsafe}
benchmark, but it also pays off for the x86-64 port, where we could reduce the
execution time of floating-point benchmarks by up to $30\%$ (compared to the
earlier \textsc{OCamlJit2} prototype).

\section{Performance} \label{section:Performance}

With the x86 port in place we were finally able to compare the performance of
\textsc{OCamlJit} \cite{Starynkevitch04} and \textsc{OCamlJit2} running on the
same machine. We measured the performance on three different systems, one x86
box for comparison with \textsc{OCamlJit}, and two x86-64 machines with
different processors and operating systems to test-drive the recent improvements
with our primary 64-bit targets:
\begin{itemize}
\item A MacBook with an Intel Core 2 Duo ``Penryn'' 2.4 GHz CPU (3 MiB L2 Cache), and
  4 GiB RAM, running Mac OS X 10.6.4. The C compiler is \texttt{gcc-4.2.1} (Apple Inc.
  build 5664).
\item A Fujitsu Siemens Primergy server with two Intel Xeon E5520 2.26GHz CPUs (8 MiB L2 Cache, 4 Cores),
  and 12 GiB RAM, running CentOS release 5.5 (Final) with Linux/x86\_64 2.6.18-194.17.1.el5.
  The C compiler is \texttt{gcc-4.2.1} (Red Hat 4.1.2-48).
\item A Fujitsu Siemens Primergy server with an Intel Pentium 4 ``Northwood'' 2.4 GHz CPU (512 KiB L2 Cache),
  and 768 MiB RAM, running Debian testing as of 2010/11/20 with Linux/i686 2.6.32-3-686.
  The C compiler is \texttt{gcc-4.4.5} (Debian 4.4.5-6).
\end{itemize}
The \textsc{OCaml} distribution used for the tests is 3.12.0. The \textsc{OCamlJit2} version
is the tagged revision \texttt{ocamljit2-2010-tr2}, compiled with a \texttt{gcc} optimization
level of \texttt{-O} (we used \texttt{-O3} in the previous measurement, but that seems to
trigger compilation bugs with recent \texttt{gcc} versions). For \textsc{OCamlJit} we had to
use \textsc{OCaml} 3.08.4, because building it with 3.12.0 caused mysterious crashes in some
test cases. We used the most recent version of \textsc{Gnu Lightning} \cite{Bonzini10} available from
the Git repository at the time of this writing (commit \texttt{d2239c223ad22a0e9d7a9909c46d2ac4e5bc0e7f}).

\begin{table*}[p]
  \footnotesize
  \centering
  \begin{tabular}{l|rrrr|rrrrrr}
    \multicolumn{1}{l|}{\large invocation}
    & \multicolumn{4}{c|}{{\large time} (cpu sec.)}
    & \multicolumn{6}{c}{\large speedup}
    \\
    command
    & $t_{byt}$ & $t_{jit}$ & $t_{jit2}$ & $t_{opt}$
    & $\sigma^{jit}_{byt}$ & $\sigma^{jit2}_{byt}$ & $\sigma^{opt}_{byt}$
    & $\sigma^{jit2}_{jit}$ & $\sigma^{opt}_{jit}$ & $\sigma^{opt}_{jit2}$
    \\
    \hline
    \texttt{almabench} & $27.61$ & & $8.58$ & $4.47$ &  & $3.22$ & $6.17$ &  &  & $1.92$\\
    \texttt{almabench.unsafe} & $27.54$ &  & $6.14$ & $4.35$ &  & $4.48$ & $6.33$ &  &  & $1.41$\\
    \texttt{bdd} & $8.46$ &  & $2.00$ & $0.67$ &  & $4.23$ & $12.66$ &  &  & $2.99$\\
    \texttt{boyer} & $4.33$ &  & $1.66$ & $1.05$ &  & $2.61$ & $4.11$ &  &  & $1.57$\\
    \texttt{fft} & $5.69$ &  & $1.98$ & $0.64$ &  & $2.88$ & $8.96$ &  &  & $3.11$\\
    \texttt{nucleic} & $14.77$ &  & $3.24$ & $0.80$ &  & $4.56$ & $18.53$ &  &  & $4.06$\\
    \texttt{quicksort} & $6.78$ &  & $1.28$ & $0.23$ &  & $5.31$ & $29.22$ &  &  & $5.50$\\
    \texttt{quicksort.unsafe} & $4.07$ &  & $0.84$ & $0.19$ &  & $4.86$ & $21.07$ &  &  & $4.34$\\
    \texttt{soli} & $0.17$ &  & $0.04$ & $0.01$ &  & $4.81$ & $17.30$ &  &  & $3.60$\\
    \texttt{soli.unsafe} & $0.14$ &  & $0.02$ & $0.01$ &  & $6.85$ & $17.12$ &  &  & $2.50$\\
    \texttt{sorts} & $19.42$ &  & $7.24$ & $3.71$ &  & $2.68$ & $5.23$ &  &  & $1.95$\\
  \end{tabular}
  \caption{Running time and speedup (Intel Core 2 Duo, Mac OS X 10.6)}
  \label{table:Running_time_and_speedup_Intel_Core_2_Duo}
\end{table*}

\begin{table*}[p]
  \footnotesize
  \centering
  \begin{tabular}{l|rrrr|rrrrrr}
    \multicolumn{1}{l|}{\large invocation}
    & \multicolumn{4}{c|}{{\large time} (cpu sec.)}
    & \multicolumn{6}{c}{\large speedup}
    \\
    command
    & $t_{byt}$ & $t_{jit}$ & $t_{jit2}$ & $t_{opt}$
    & $\sigma^{jit}_{byt}$ & $\sigma^{jit2}_{byt}$ & $\sigma^{opt}_{byt}$
    & $\sigma^{jit2}_{jit}$ & $\sigma^{opt}_{jit}$ & $\sigma^{opt}_{jit2}$
    \\
    \hline
    \texttt{almabench} & $28.52$ &  & $9.40$ & $5.36$ &  & $3.03$ & $5.32$ &  &  & $1.76$\\
    \texttt{almabench.unsafe} & $26.52$ &  & $7.87$ & $5.56$ &  & $3.37$ & $4.77$ &  &  & $1.42$\\
    \texttt{bdd} & $9.51$ &  & $2.03$ & $0.59$ &  & $4.69$ & $16.23$ &  &  & $3.46$\\
    \texttt{boyer} & $3.77$ &  & $1.68$ & $1.02$ &  & $2.24$ & $3.70$ &  &  & $1.65$\\
    \texttt{fft} & $5.65$ &  & $1.47$ & $0.34$ &  & $3.85$ & $16.62$ &  &  & $4.32$\\
    \texttt{nucleic} & $14.00$ &  & $3.28$ & $0.78$ &  & $4.27$ & $17.86$ &  &  & $4.18$\\
    \texttt{quicksort} & $5.32$ &  & $1.26$ & $0.23$ &  & $4.23$ & $22.81$ &  &  & $5.39$\\
    \texttt{quicksort.unsafe} & $5.48$ &  & $0.88$ & $0.18$ &  & $6.25$ & $29.80$ &  &  & $4.77$\\
    \texttt{soli} & $0.14$ &  & $0.03$ & $0.01$ &  & $4.18$ & $15.33$ &  &  & $3.67$\\
    \texttt{soli.unsafe} & $0.12$ &  & $0.02$ & $0.01$ &  & $6.56$ & $16.86$ &  &  & $2.57$\\
    \texttt{sorts} & $21.50$ &  & $7.08$ & $3.61$ &  & $3.03$ & $5.96$ &  &  & $1.96$\\
  \end{tabular}
  \caption{Running time and speedup (Intel Xeon, CentOS 5.5)}
  \label{table:Running_time_and_speedup_Intel_Xeon}
\end{table*}

\begin{table*}[p]
  \footnotesize
  \centering
  \begin{tabular}{l|rrrr|rrrrrr}
    \multicolumn{1}{l|}{\large invocation}
    & \multicolumn{4}{c|}{{\large time} (cpu sec.)}
    & \multicolumn{6}{c}{\large speedup}
    \\
    command
    & $t_{byt}$ & $t_{jit}$ & $t_{jit2}$ & $t_{opt}$
    & $\sigma^{jit}_{byt}$ & $\sigma^{jit2}_{byt}$ & $\sigma^{opt}_{byt}$
    & $\sigma^{jit2}_{jit}$ & $\sigma^{opt}_{jit}$ & $\sigma^{opt}_{jit2}$
    \\
    \hline
    \texttt{almabench} & $57.72$ & $28.81$ & $20.09$ & $17.20$ & $2.00$ & $2.87$ & $3.35$ & $1.43$ & $1.67$ & $1.17$\\
    \texttt{almabench.unsafe} & $56.08$ & $26.04$ & $16.89$ & $19.70$ & $2.15$ & $3.32$ & $2.85$ & $1.54$ & $1.32$ & $0.86$\\
    \texttt{bdd} & $15.51$ & $6.25$ & $4.90$ & $1.14$ & $2.48$ & $3.17$ & $13.56$ & $1.28$ & $5.47$ & $4.28$\\
    \texttt{boyer} & $8.31$ & $4.30$ & $3.84$ & $1.96$ & $1.93$ & $2.16$ & $4.23$ & $1.12$ & $2.19$ & $1.96$\\
    \texttt{fft} & $13.70$ & $7.20$ & $5.23$ & $3.27$ & $1.90$ & $2.62$ & $4.19$ & $1.38$ & $2.20$ & $1.60$\\
    \texttt{nucleic} & $33.21$ & $14.65$ & $7.56$ & $2.11$ & $2.27$ & $4.39$ & $15.72$ & $1.94$ & $6.94$ & $3.58$\\
    \texttt{quicksort} & $10.93$ & $3.64$ & $3.10$ & $0.34$ & $3.00$ & $3.53$ & $32.14$ & $1.18$ & $10.72$ & $9.11$\\
    \texttt{quicksort.unsafe} & $9.15$ & $2.79$ & $2.16$ & $0.28$ & $3.28$ & $4.23$ & $32.67$ & $1.29$ & $9.96$ & $7.73$\\
    \texttt{soli} & $0.32$ & $0.08$ & $0.06$ & $0.02$ & $3.86$ & $5.40$ & $20.25$ & $1.40$ & $5.25$ & $3.75$\\
    \texttt{soli.unsafe} & $0.27$ & $0.06$ & $0.03$ & $0.01$ & $4.79$ & $9.57$ & $22.33$ & $2.00$ & $4.67$ & $2.33$\\
    \texttt{sorts} & $47.18$ & $20.62$ & $16.55$ & $6.26$ & $2.29$ & $2.85$ & $7.53$ & $1.25$ & $3.29$ & $2.64$\\
  \end{tabular}
  \caption{Running time and speedup (Intel Pentium 4, Debian testing)}
  \label{table:Running_time_and_speedup_Intel_Pentium_4}
\end{table*}

The benchmark programs used to measure the performance are the following test programs from the
\texttt{testsuite/tests} folder of the \textsc{OCaml} 3.12.0 distribution:
\begin{itemize}
\item \texttt{almabench} is a number-crunching benchmark designed for cross-language
  comparisons. \texttt{almabench.unsafe} is the same program compiled with the \texttt{-unsafe}
  compiler switch.
\item \texttt{bdd} is an implementation of binary decision diagrams, and therefore a good
  test for the symbolic computation performance.
\item \texttt{boyer} is a term manipulation benchmark.
\item \texttt{fft} is an implementation of the Fast Fourier Transformation.
\item \texttt{nucleic} is another floating-point benchmark.
\item \texttt{quicksort} is an implementation of the well-known QuickSort algorithm
  on arrays and serves as a good test for loops.
\item \texttt{soli} is a simple solitaire solver, well suited for testing the performance
  of non-trivial, short running programs.
\item \texttt{sorts} is a test bench for various sorting algorithms.
\end{itemize}
For our tests we measured the total execution time of the benchmarks, given as combined
system and user CPU time, in seconds.
Table~\ref{table:Running_time_and_speedup_Intel_Core_2_Duo}
lists the running times and speedups for the Intel Core 2 Duo,
table~\ref{table:Running_time_and_speedup_Intel_Xeon}
for the Intel Xeon,
and table~\ref{table:Running_time_and_speedup_Intel_Pentium_4}
for the Intel Pentium 4. We compare the byte-code interpreter time $t_{byt}$ to
the \textsc{OCamlJit} time $t_{jit}$ (if available), the \textsc{OCamlJit2} time
$t_{jit2}$, and the time $t_{opt}$ taken by the same program compiled with the
optimizing native code compiler \texttt{ocamlopt}. The tables also list the relative
speedups
$\sigma^{jit}_{byt} = \frac{t_{byt}}{t_{jit}}$,
$\sigma^{jit2}_{byt} = \frac{t_{byt}}{t_{jit2}}$,
$\sigma^{opt}_{byt} = \frac{t_{byt}}{t_{opt}}$,
$\sigma^{jit2}_{jit} = \frac{t_{jit}}{t_{jit2}}$,
$\sigma^{opt}_{jit} = \frac{t_{jit}}{t_{opt}}$, and
$\sigma^{opt}_{jit2} = \frac{t_{jit2}}{t_{opt}}$, where bigger values are better. The
times were collected by executing each benchmark $5$ times with every engine, and
using the timings of the fastest run.

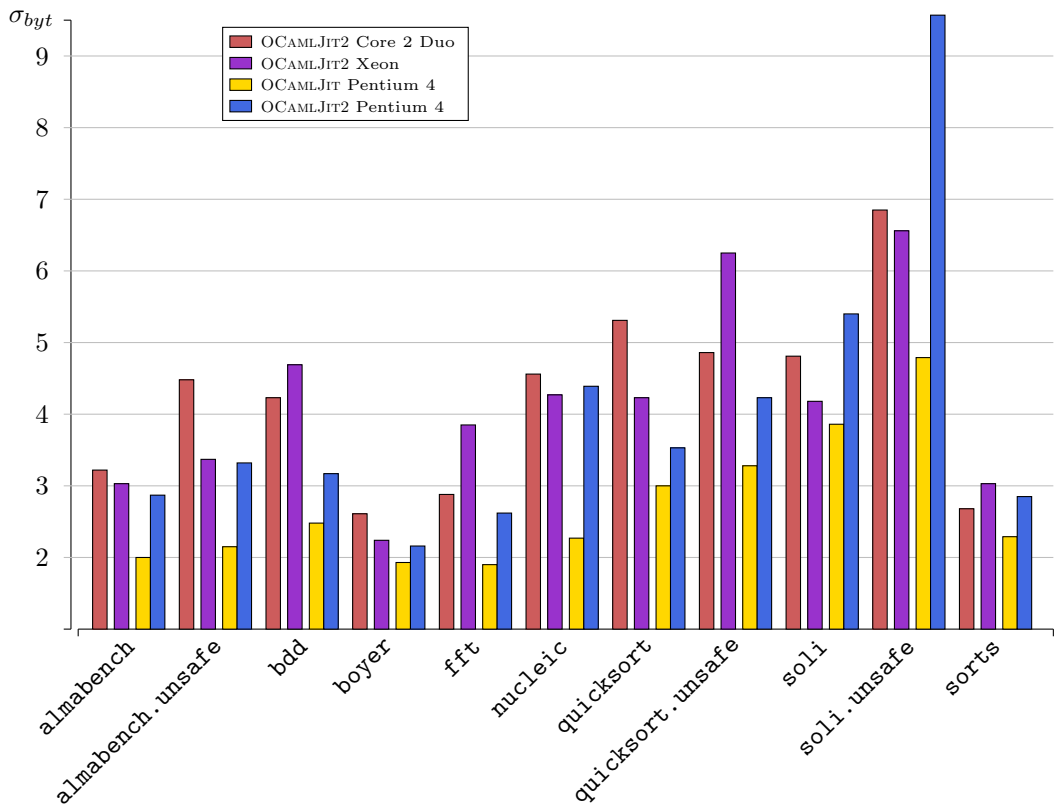
\begin{figure*}[htb]
  \centering
  \footnotesize
  \begin{tikzpicture}[scale=.95]
    \draw (0cm,0cm) -- (13.5cm,0cm);  
    \draw (0cm,0cm) -- (0cm,-0.1cm);  
    \draw (13.5cm,0cm) -- (13.5cm,-0.1cm);  
    \draw (-0.1cm,0cm) -- (-0.1cm,8.5cm);  
    \draw (-0.1cm,0cm) -- (-0.2cm,0cm);  
    \draw (-0.1cm,8.5cm) -- (-0.2cm,8.5cm) node [left] {$\sigma_{byt}$};  
    \foreach \x in {2,...,9}
    {
      \draw[gray!50, text=black] (-0.2 cm,\x cm - 1cm) -- (13.5 cm,\x cm - 1cm) node at (-0.5 cm,\x cm - 1cm) {\x};
    }
    \foreach \name/\x/\a/\b/\c/\d in
    {
      almabench/0 / 3.22 / 3.03 / 2.00/2.87,
      almabench.unsafe/1 / 4.48 / 3.37 / 2.15/3.32,
      bdd/2 / 4.23 / 4.69 / 2.48/3.17,
      boyer/3 / 2.61 / 2.24 / 1.93/2.16,
      fft/4 / 2.88 / 3.85 / 1.90/2.62,
      nucleic/5 / 4.56 / 4.27 / 2.27/4.39,
      quicksort/6 / 5.31 / 4.23 / 3.00/3.53,
      quicksort.unsafe/7 / 4.86 / 6.25 / 3.28/4.23,
      soli/8 / 4.81 / 4.18 /  3.86/5.40,
      soli.unsafe/9 / 6.85 / 6.56 / 4.79/9.57,
      sorts/10 / 2.68 / 3.03 / 2.29/2.85
    }
    {
      \pgfmathsetmacro\xinc{0.2}
      \pgfmathsetmacro\xoff{\x * 1.2 + \xinc}
      \pgfmathsetmacro\aoff{\a - 1.0}
      \pgfmathsetmacro\boff{\b - 1.0}
      \pgfmathsetmacro\coff{\c - 1.0}
      \pgfmathsetmacro\doff{\d - 1.0}
      \draw[fill=core2jit2] (\xoff cm + 0 * \xinc cm, 0 cm) rectangle (1 * \xinc cm + \xoff cm, \aoff);
      \draw[fill=xeonjit2] (\xoff cm + 1 * \xinc cm + .1cm, 0 cm) rectangle (2 * \xinc cm + \xoff cm + .1cm, \boff);
      \draw[fill=p4jit] (\xoff cm + 2 * \xinc cm + .2cm, 0 cm) rectangle (3 * \xinc cm + \xoff cm + .2cm, \coff);
      \draw[fill=p4jit2] (\xoff cm + 3 * \xinc cm + .2cm, 0 cm) rectangle (4 * \xinc cm + \xoff cm + .2cm, \doff);
      \node[rotate=45,left] at (\xoff cm + 3 * \xinc cm, -.1cm) {\texttt{\name}};
    };
    \filldraw[fill=white] (2,8.4) rectangle (5.4,7.1);
    \foreach \engine/\proc/\color/\y in
    {
      OCamlJit2/Core 2 Duo/core2jit2/0,
      OCamlJit2/Xeon/xeonjit2/1,
      OCamlJit/Pentium 4/p4jit/2,
      OCamlJit2/Pentium 4/p4jit2/3
    }
    {
      \pgfmathsetmacro\xoff{2.1}
      \pgfmathsetmacro\yoff{8.3 - \y * 0.3}
      \filldraw[fill=\color] (\xoff,\yoff-0.02) rectangle (\xoff+0.3,\yoff-0.18)
        node at (\xoff+0.3,\yoff-0.1) [right] {\tiny\textsc{\engine} \proc};
    };
  \end{tikzpicture}
  \caption{Speedup relative to the byte-code interpreter}
  \label{figure:Speedup_relative_to_the_byte_code_interpreter}
\end{figure*}

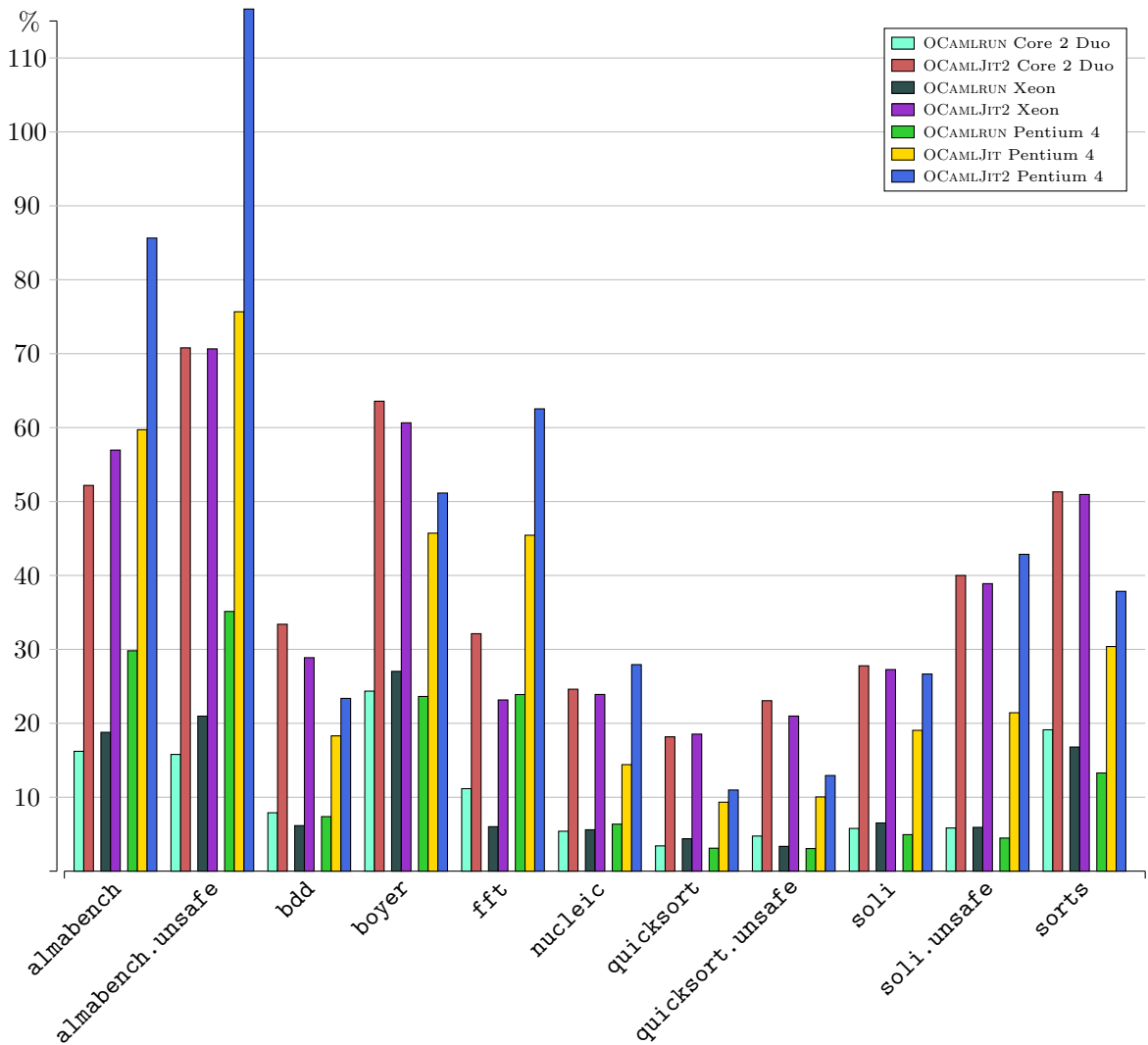
\begin{figure*}[htb]
  \centering
  \footnotesize
  \begin{tikzpicture}
    \definecolor{color}{HTML}{483D8B}
    \draw (0cm,0cm) -- (14.5cm,0cm);  
    \draw (0cm,0cm) -- (0cm,-0.1cm);  
    \draw (14.5cm,0cm) -- (14.5cm,-0.1cm);  
    \draw (-0.1cm,0cm) -- (-0.1cm,11.5cm);  
    \draw (-0.1cm,0cm) -- (-0.2cm,0cm);  
    \draw (-0.1cm,11.5cm) -- (-0.2cm,11.5cm) node [left] {\%};  
    \foreach \y in {1,...,11}  
    {
      \pgfmathtruncatemacro\ytext{\y * 10}
      \draw[gray!50, text=black] (-0.2 cm,\y cm) -- (14.5 cm,\y cm)
        node at (-0.5 cm,\y cm) {\ytext};  
    };
    \foreach \name/\x/\a/\b/\c/\d/\e/\f/\g in
    {
      almabench/0 / 1.621/5.218 / 1.878/5.696 / 2.981/5.971/8.564,
      almabench.unsafe/1 / 1.579/7.079 / 2.097/7.065 / 3.513/7.567/11.662,
      bdd/2 / 0.790/3.340 / 0.616/2.888 / 0.738/1.830/2.337,
      boyer/3 / 2.435/6.357 / 2.703/6.065 / 2.363/4.572/5.115,
      fft/4 / 1.116/3.212 / 0.602/2.316 / 2.389/4.544/6.254,
      nucleic/5 / 0.540/2.461 / 0.560/2.390 / 0.636/1.441/2.794,
      quicksort/6 / 0.342/1.817 / 0.438/1.854 / 0.311/0.933/1.098,
      quicksort.unsafe/7 / 0.475/2.306 / 0.336/2.098 / 0.306/1.004/1.294,
      soli/8 / 0.578/2.778 / 0.652/2.727 / 0.494/1.905/2.667,
      soli.unsafe/9 / 0.584/4.000 / 0.593/3.889 / 0.448/2.143/4.286,
      sorts/10 / 1.912/5.131 / 1.679/5.095 / 1.328/3.038/3.785
    }
    {
      \pgfmathsetmacro\xinc{0.13}
      \pgfmathsetmacro\xoff{\x * 1.3 + \xinc}
      \draw[fill=core2byt] (\xoff cm + 0 * \xinc cm, 0 cm) rectangle (1 * \xinc cm + \xoff cm, \a);
      \draw[fill=core2jit2] (\xoff cm + 1 * \xinc cm, 0 cm) rectangle (2 * \xinc cm + \xoff cm, \b);
      \draw[fill=xeonbyt] (\xoff cm + 2 * \xinc cm + .1cm, 0 cm) rectangle (3 * \xinc cm + \xoff cm + .1cm, \c);
      \draw[fill=xeonjit2] (\xoff cm + 3 * \xinc cm + .1cm, 0 cm) rectangle (4 * \xinc cm + \xoff cm + .1cm, \d);
      \draw[fill=p4byt] (\xoff cm + 4 * \xinc cm + .2cm, 0 cm) rectangle (5 * \xinc cm + \xoff cm + .2cm, \e);
      \draw[fill=p4jit] (\xoff cm + 5 * \xinc cm + .2cm, 0 cm) rectangle (6 * \xinc cm + \xoff cm + .2cm, \f);
      \draw[fill=p4jit2] (\xoff cm + 6 * \xinc cm + .2cm, 0 cm) rectangle (7 * \xinc cm + \xoff cm + .2cm, \g);
      \node[rotate=45,left] at (\xoff cm + 5 * \xinc cm, -.1cm) {\texttt{\name}};
    };
    \filldraw[fill=white] (11,11.4) rectangle (14.25,9.2);
    \foreach \engine/\proc/\color/\y in
    {
      OCamlrun/Core 2 Duo/core2byt/0,
      OCamlJit2/Core 2 Duo/core2jit2/1,
      OCamlrun/Xeon/xeonbyt/2,
      OCamlJit2/Xeon/xeonjit2/3,
      OCamlrun/Pentium 4/p4byt/4,
      OCamlJit/Pentium 4/p4jit/5,
      OCamlJit2/Pentium 4/p4jit2/6
    }
    {
      \pgfmathsetmacro\xoff{11.1}
      \pgfmathsetmacro\yoff{11.3 - \y * 0.3}
      \filldraw[fill=\color] (\xoff,\yoff-0.02) rectangle (\xoff+0.3,\yoff-0.18)
        node at (\xoff+0.3,\yoff-0.1) [right] {\tiny\textsc{\engine} \proc};
    };
  \end{tikzpicture}
  \caption{Performance relative to \texttt{ocamlopt}}
  \label{figure:Performance_relative_to_ocamlopt}
\end{figure*}

Figure~\ref{figure:Speedup_relative_to_the_byte_code_interpreter} highlights the speedup of
the JIT engines relative to the byte-code interpreter on the three test systems. The x86-64
port received another nice speedup in the floating point benchmarks compared to the earlier
prototype due to the improvements described in section~\ref{subsection:Floating_point_optimizations}.
On x86 \textsc{OCamlJit2} provides a performance boost of $2.2$ to $9.6$ compared to the
byte-code interpreter, and is $1.1$ to $2.0$ times faster than \textsc{OCamlJit}. This is
especially noticable with short running programs like \texttt{soli.unsafe}, where \textsc{OCamlJit2}
benefits from the reduced compilation overhead, and in floating-point programs, where \textsc{OCamlJit}
wins because of clever floating point optimizations and its use of the SSE2 extension.

Figure~\ref{figure:Performance_relative_to_ocamlopt} shows the performance of the byte-code
interpreter and the JIT engines relative to the optimizing native code compiler \texttt{ocamlopt}.
One rather surprising result was the bad performance of the generated x86 native code for
the \texttt{almabench.unsafe} benchmark, where \textsc{OCamlJit2} was able to beat the
native code compiler by a factor of $1.2$. This may be related to the use of SSE2 instead of
x87 instructions, which are generally faster, but it also seems that we have spotted a problem
within the x86 port of \texttt{ocamlopt} here (unfortunately, we were unable to track down
the issue). On a related note, we have also spotted some issues with the
x86-64 floating-point code generated by \texttt{ocamlopt} and already submitted an appropriate
patch\footnote{\url{http://caml.inria.fr/mantis/view.php?id=5180}}, which will be available in
\textsc{OCaml} 3.12.1, yielding performance improvements of $6-13\%$ with floating-point programs.

In general, floating-point benchmarks benefit a lot when used with \textsc{OCamlJit2},
which was somewhat expected, in particular with the optimizations implemented in the latest
prototype. \textsc{OCamlJit} does a respectable job, but is unable to compete with
\textsc{OCamlJit2} performance-wise. This is probably caused by the better compilation
scheme used by \textsc{OCamlJit2} and also related to the use of \textsc{Gnu Lightning}
within \textsc{OCamlJit}. Nevertheless, it is this use of \textsc{Gnu Lightning} which
makes \textsc{OCamlJit} slightly more portable than \textsc{OCamlJit2} (three supported
platforms in case of \textsc{OCamlJit}, compared to only two with \textsc{OCamlJit2}).

\section{Conclusions and further work} \label{section:Conclusions_and_further_work}

Our results show that Just-In-Time compilation of \textsc{OCaml} byte-code can lead
to some significant speedup (at least twice as fast the byte-code interpreter in all
benchmarks). Starting out with a simple compilation scheme and applying some clever
optimizations led to impressive performance gains. But we have also noticed that our
approach is somewhat limited, which is mostly related to the nature of the \textsc{OCaml}
byte-code and the design choices made for the interpreter runtime.
The \textsc{OCaml} byte-code and runtime were certainly designed with ``fast interpretation''
in mind \cite{Leroy90} and perform quite well in this area, but this same fact also limits the
possibilities for effective Just-In-Time compilation if one wants to avoid touching too
many aspects of the runtime. Using a register machine as used by Dalvik \cite{Bornstein08}
or Parrot \cite{Coleda10} instead of the stack machine for the byte-code virtual machine would
make it easier to apply JIT techniques -- it would in fact make Just-In-Time compilation
the natural implementation choice for the virtual machine.

Implementations of other runtimes, like the various JVMs \cite{LindholmYellin99} or the
Common Language Runtime \cite{Ecma335},
show that it is indeed possible to perform efficient JIT compilation with stack virtual machines, but
these runtimes use expensive compilation techniques for instruction selection, register
allocation and instruction scheduling, whose applicability is questionable
within the scope of the \textsc{OCaml} byte-code runtime. For example, in order to
effectively reduce the overhead of closure application and return in \textsc{OCaml}
byte-code execution, one would most likely need to perform interprocedural register allocation,
which mandates the availability of global control and data flow information, both of which are
difficult to collect efficiently.
It may indeed be possible to design a Just-In-Time compiler for the \textsc{OCaml}
byte-code, which generates code as fast as the code generated by \texttt{ocamlopt},
using standard compilation techniques \cite{Aho06}, but such a JIT engine comes at
a high cost with respect to maintainability and execution speed of the JIT compiler, and it
is questionable whether it is worth this cost, especially since there is already an optimizing native code
compiler, which limits the possible use cases for the Just-In-Time compiler.

The main application of \textsc{OCamlJit2} is the interactive top-level and other
dynamic code generation environments like \textsc{MetaOCaml} \cite{Taha06}. The
\textsc{OCaml} repository already contains an experimental ``native top-level''
\texttt{ocamlnat} for this purpose, which uses the functionality of \texttt{ocamlopt}
to generate efficient native code at runtime and execute it via the native code
runtime. Improving \texttt{ocamlnat} may provide a better way to gain an efficient
top-level, and we are currently evaluating what would need to be done in this
area.

\section*{Acknowledgements}

We would like to thank Xavier Leroy and the \textsc{OCaml} community for their
useful comments on the earlier prototype, as well as Christian Uhrhan and
Mehrnush Afshar for their careful proof-reading.

\bibliographystyle{abbrv}
\bibliography{citations}

\end{document}